\documentstyle{article}
\begin{document}

\vspace*{3mm}

\begin{center}

{\LARGE \bf 4-D Quantum Dilaton Gravity \vspace*{2mm}\\
During Inflation \vspace*{4mm}\\
and Renormalization at One loop }

\vspace{15mm}

{\sc Hiroyuki Takata}
\footnote{ E-MAIL: takata@theory.kek.jp,
 URL: http://theory.kek.jp/~takata/}\\
National Laboratory for High Energy Physics (KEK), 
Tsukuba, Ibaraki 305, Japan.\\
    and\\
Department of Physics, Hiroshima University,  
Higashi-Hiroshima 739, Japan. \\
\vskip 5mm

\vspace{15mm}

{\bf Abstract}

\end{center}

\baselineskip 8mm
We consider 4D quantum-dilaton gravity with the most general
 coupling in a
 homogeneous and isotropic universe, especially an inflationary one,
 which
 is essentially characterized by an exponentially expanding scale
 factor
 with time.
We show that on the inflationary background this theory can be
 miraculously
 renormalized, at least at the one-loop level, which must be an
 effective
 theory during the inflation of the un-constructed complete quantum
 theory
 of gravity.  
\section{Introduction and Summary}
\hspace*{5mm}
At present, no complete unified theory which includes quantum
 gravity has
 yet been constructed. However if we believe in the existence of
 such a
 theory, and wish to know the history of our universe controlled by
 this
 theory, we must find it as soon as possible.  
Our point of view is that such a fundamental theory of quantum
 gravity is
 either string theory or another theory like it. 
We do not place any restrictions on which is true.
Furthermore, we do not remove the possibility of quantum gravity to
 be a
 local field theory with renormalizability. 
In Any case there is an effective theory of fundamental quantum
 gravity for
 low energy or special background.

A candidate for a consistent theory of quantized gravity is string
 theory.
A low-energy effective theory of a string below the Plank scale
 represented
 by a metric and a dilaton is well known \cite{GSW}. 
Such an effective action arises in the form of a power-series type
 of a
 slope parameter ($\alpha^{\prime}$); the standard point of view is
 that
 the higher orders in such an expansion correspond to higher
 energies.
From this point of view, at a lower energy scale the action for
 gravity has
 the form of a lower derivative dilaton action.
The Einstein gravity coupled to scalars is nonrenormalizable as
 naive power
 counting, and higher derivative gravity is a renormalizable
 \cite{St};
 however, it is not unitary within a perturbation scheme
 \cite{BOS}.
Of course, it is not strange that a useful local field theory of
 gravity
 covering all energy regions does not exist, though its possibility
 cannot
 be rejected.
It is important to know whether a renormalizable local field theory
 of
 gravity constructed by metric exists or not, and what type of environment
 would allow its existence.
It is well known \cite{We} that the standard cosmological model works well
 for the red shift, cosmic microwave background, and nucleosynthesis.
However, its naive version cannot explain some problems: flatness and
 horizon.
An inflationary model was proposed by Guth \cite{Gu} in order to settle
 these problems, which was implemented by means of a first-order phase
 transition.
Unfortunately, it cannot be a realistic model, as Guth himself has pointed
 out. After a time, other models were proposed with inflation as a second
 order phase transition \cite{Li1} or ``chaotic'' inflation \cite{Li2},
 which have fine-tuned parameters that depend on fundamental theory before
 the inflation begins.
The inflation model can be well explained  within the framework of the
 grand unified theory (GUT) of particles at least qualitatively
 \cite{Gu,Li1}.
In the GUT model all interactions, except for gravity, may be unified on
  a scale (GUT scale) below the Plank scale, where the gauge symmetry of
 GUT spontaneously breaks down to one of the standard models of elementary
 particles with some field which has a non-zero vacuum expectation value.
Naively, it seems that this field can be identified with the field driving
 inflation. However, quantitatively such a model cannot be constructed well
 while being  consistent with observational data concerning the
  inhomogeneity of the microwave background radiation \cite{ST}.
Thus, in some model the inflationary phase transition is supposed to begin
 above the  GUT scale, say, the Plank scale \cite{Li2,Li3}.
At this point, a  reliable discussion of inflation can be realized only
 after a quantum theory of gravity is constructed. 
For instance, within the framework of the string theory the cosmology near
 to the Plank scale has been discussed \cite{Ve1,Ve2,Ve3,TV,Ts2,Ts3};
 furthermore, the relation between the inflationary universe phase to usual
 A Freedman-universe phase is considered \cite{Ve1,Ve3,Ts2} using the fact
 that  scale factor duality \cite{Ve2,Ts1} is  a subset of the T duality
 \cite{KY,Sc}, where the dilaton is identified as being  the field driving
 inflation.\\
Our stand point is that this dilaton drives inflation, however, since our
 action is more general than a low-energy effective action of a string we
 do not restrict our scalar field to the usual dilaton in string theory.
We consider an arbitrary action with a metric and a scalar field called
 ``dilaton'' having  two derivatives, and consider the inflation driven by
 the dilaton near to the Plank scale where quantum effect of gravity is
 very important. 

Until quite recently, from long time ago, several studies were performed to
 calculate the divergence of effective action of four-dimensional gravity
 \cite{HV,CD,BKK,ST1}. 
In the pure Einstein action case without a cosmological term, it was
 originally calculated at the one-loop level by t'Hooft and Veltman
 \cite{HV}. 
They found that the action is not renormalizable off mass shell, but is
 finite on mass shell at the one-loop level.
Furthermore, although the pure Einstein action with a cosmological constant
 is renormalizable \cite{CD}, if one introduces matter fields the one loop
 renormalizability is lost, even on mass shell.
Recently \cite{ST1,ST2}, in the action\footnote{In Ref \cite{ST1,ST2}, we
 used a convention $C(\phi)=-2 B(\phi)\Lambda(\phi)$}
\begin{equation}
S= \int d^4x \sqrt{-g}\; \{ A(\phi)g^{\mu\nu}\partial_{\mu}\phi
\partial_{\nu}\phi + B(\phi)R -2B(\phi) \Lambda(\phi) \}
\label{action}
\end{equation}
we considered the divergence of the effective action, which is the most
 general class with less than two derivatives for a scalar and a metric,
 while explicitly leaving functions $A,B,\Lambda$ arbitrary.
Classically, theory (\ref{action}) can be classified into two cases
 according to whether they have conformal symmetry or not.
In the later case, on an arbitrary back-ground space-time we found models
 which are finite in the case without a cosmological term, and are
 renormalizable in the case with it by fine-tuning of functional form of
 $A(\phi), B(\phi), \Lambda(\phi)$ at the one loop level on mass shell.
This property is the same as pure Einstein gravity with or without a
 cosmological term, although there is a matter coupling to the metric in
 our model.
On the other hand, an analysis of quantum theory was considered
 \cite{Ts3,IMM,AIT} on maximal symmetric space-time where a number of
 Killing vectors exist.
It includes De Sitter space-time, which is a limiting case of the usual
 inflationary universe which has an exponentially expanding scale factor. 

In the present paper we consider the action (\ref{action}) on a homogeneous
 and isotropic four-dimensional background space-time (spatially maximal
 symmetric space-time), and calculate the divergence of the effective
 action.
We especially consider the divergence on inflationary background space-time
 (maximal symmetric one), and found that one-loop divergence can be
 renormalized. 
With a  special choice of $A(\phi),B(\phi),\Lambda(\phi)$, (\ref{action})
 is equivalent to $R^2$ gravity, by which we show that this case is
 renormalizable, and consider a renormalization group analysis on the
 inflationary back-ground \cite{ST3}.
    
This paper is organized as follows.
In section 2,  
considering model (\ref{action}) in a homogeneous and isotropic
 four-dimensional background space-time, we found that the classical
 equations of motion of the metric are the  ``Hubble'' equation and the
 energy momentum-conservation law, that the equation of motion of the
 dilaton ($\phi$) means the state of universe being defined by a  ``state''
 equation between the energy density and the pressure driven by dilaton,
 and that there is a maximal symmetric solution as the simplest one. 
In section 3 
we discuss our  calculation of the divergence of effective action of
 (\ref{action}) on homogeneous and isotropic four dimensional background
 space-time using the back-ground field method \cite{Ab} and the
 Schwinger-DeWitt technique \cite{De,BOS}.
Next, we show that the structure of the divergence on maximal symmetric
 space-time is miraculously simple, and renormalizable, at least at the
 one-loop level.
\section{Analysis in Classical level}
\hspace*{5mm}
We consider gravity with the general coupling to a scalar in which the
 action is (\ref{action})\footnote{In this paper we restrict 
$B \neq 0$ and $A \neq \frac{3}{2}\frac{B_1^2}{B}$, 
where we write $X_n:=\frac{d^n X(\phi)}{d \phi^n}$ for any function
 $X(\phi)$}.
For a special classical back-ground it may be  realized in the very early
 universe.
First, in this section we analyze this theory at the classical level. 
\subsection{Classical Equation of Motion }
\hspace*{5mm}
The classical equations of motion for $g_{\mu\nu}$ and $\phi$ are
\begin{equation}
R_{\mu\nu}-\frac{1}{2}R g_{\mu\nu}+ \Lambda g_{\mu\nu}=T_{\mu\nu} 
\;\;\;\;\;\;(\;\mbox{for}\;g_{\mu\nu})
\end{equation}
and 
\begin{equation}
 B_1 R -2(B\Lambda)_1 -A_1(\nabla \phi)^2 -2A (\Box \phi) =0 
\;\;\;\;\;\;(\;\mbox{for}\;\phi)\;,
\end{equation}
where 
\[
T_{\mu\nu}:=
\]
\begin{equation}
-\left(\left( \frac{B_2}{B}-\frac{A}{2 B}\right)(\nabla \phi)^2
+ \frac{B_1}{B}(\Box \phi)\right)g_{\mu\nu}
+\left( \frac{B_2}{B}-\frac{A}{ B}\right)(\nabla_{\mu} \phi)
(\nabla_{\nu} \phi)+\frac{B_1}{B}(\nabla_{\mu}\nabla_{\nu}\phi)\;.
\end{equation}
In the next subsection we consider a homogeneous and isotropic universe
 (maximally symmetric three-dimensional space) with the Robertson-Walker
 metric, which is a realistic model of a hot universe. 
Also in the following subsection we consider a maximally symmetric
 space-time, which is the simplest model of the universe during inflation.

\subsection{Maximally  Symmetric Space}
\hspace*{5mm}
It is well known that in D-dimensional space-time there are a maximum of
 $\frac{D(D+1)}{2}$ Killing vectors.
That space-time is called a maximally symmetric space-time.
It is specified by \cite{We}
\begin{equation}
g_{\mu\nu}=\eta_{\mu\nu}
+ \frac{K x_\mu x_\nu }{1-K x_{\lambda}x^{\lambda}}\;,
\;\;\;\;\eta_=\mbox{diag}(\pm1,\pm1,\cdots)\;,
\;\;\;K \mbox{ is constant and } x_{\mu}:=\eta_{\mu\lambda}x^{\lambda}\;.
\end{equation}
Thus,\footnote{In this paper we use the convention 
$\Gamma^{\rho}_{\mu\nu}
=\frac{1}{2}g^{\rho \sigma}\left(g_{\mu \sigma, \nu}
 + g_{\nu \sigma, \mu} - g_{\mu \nu, \sigma} \right)$,
 $R^{\rho}_{\; \sigma \mu \nu}=\Gamma^{\rho}_{\sigma \nu ,\mu}
 +\Gamma^{\rho}_{\mu \lambda}\Gamma^{\lambda}_{\sigma \nu}
 -(\mu \leftrightarrow \nu)$,
 $R_{\mu\nu}=R^{\rho}_{\;\mu \rho \nu}$}
\begin{equation}
\Gamma ^{\mu}_{\nu\lambda}=K x^{\mu}g_{\nu\lambda}\;,
\end{equation}
\begin{equation}
R_{\mu\nu\alpha\beta}
=K \left(g_{\mu\alpha}g_{\nu\beta}-g_{\nu\alpha}g_{\mu\beta}\right)\;,
\end{equation}
\begin{equation}
R_{\mu\nu}=(D-1)K g_{\mu\nu}\;,
\end{equation}
and
\begin{equation}
R=D(D-1)K\;.
\end{equation}
Furthermore, if $K$ and the signature of $g_{\mu\nu}$ are fixed,
 any metric with a maximal symmetry can be transformed to above the
 formula by some general coordinate transformation.
In the next subsection we consider $D=3$
 and $\eta=\mbox{diag}(-1, -1, -1)$ case;  in the subsequent subsection we
 consider the $D=4$ and $\eta=\mbox{diag}(+1, -1, -1, -1)$ case.

\subsection{Homogeneous and Isotropic Universe}
\hspace*{5mm}
A homogeneous and isotropic universe is a spatially maximally symmetric
  subspace of the whole four-dimensional space-time.
In this space-time the Robertson-Walker metric is convenient,
 which is 
\begin{equation}
ds^2=dt^2 - a(t)^2\left(\frac{1}{1-k r^2}dr^2 + r^2 d\Omega^2 \right)\;,
\end{equation}
where $a(t)$ is a scale factor  and $k=\pm 1, 0$. Then,
\begin{equation}
R_{i j k l}=\kappa_t(g_{i k}g_{j l}-g_{i l}g_{j k})
,\;\;\;\;R_{0 j 0 l}=-\kappa_sg_{j l}\;,
\end{equation}
\begin{equation}
R_{\mu\nu}=\left(
\begin{array}{cc}
3 \kappa_s  & 0 \\
0 & \left( \kappa_s  + 2 \kappa_t \right)g_{i j}  
\end{array}
\right)\;,
\end{equation}
and 
\begin{equation}
R=6\left( \kappa_s  + \kappa_t \right)\;,
\end{equation}
where $\kappa_s(t) :=\frac{\ddot{a}}{a}$,
 $\kappa_t(t):=\left(\frac{\dot{a}}{a}\right)^2 + \frac{k}{a^2}$,
 and a dot means a time derivative.
In a maximally symmetric three-dimensional space, a rank-2 tensor is
 proportional to a metric ($g_{i j}$) with a coefficient which depends only
 on time, and a scalar depends only on time. Therefore, the energy momentum
 tensor ($T_{\mu\nu}$) is very simple:
\begin{equation}
T_{\mu\nu}=:\left(
\begin{array}{cc}
-\rho(t) & 0 \\
0 & p(t) g_{i j}  
\end{array}
\right)\;.
\end{equation}
The equations of motion of $g_{\mu\nu}$ and $\phi$ are  equivalent to the
 following  equations:\\
(i) ``Hubble'' equation:
\begin{equation}
\left(\frac{\dot{a}}{a}\right)^2 + \frac{k}{a^2}=\frac{\rho +\Lambda}{3}
\end{equation}
(ii) Energy conservation low:
\begin{equation}
\dot{\rho}+\dot{\Lambda}=-3 \frac{\dot{a}}{a}(\rho + p)
\end{equation}
(iii) ``State'' equation:
\begin{equation}
\left( B_1^2+\frac{2A_1 B B_1}{A}+2B(A-2B_2) \right)\rho 
-\left( 3B_1^2+2A B \right)p 
+2B_1 \left(B_1 \Lambda -B \Lambda_1 \right)=0
\end{equation}
(iv) Relation between $\rho\;,\;p$ and $\phi$:
\[
\rho =-\frac{A}{2B} \dot{\phi}^2,
\]
\begin{equation}
p=\left(\frac{B_2}{B}- \frac{A}{2B} \right)\dot{\phi}^2
 +\frac{B_1}{B}\ddot{\phi}
.
\end{equation}
\subsection{Inflationary Universe}
\hspace*{5mm}
The essence of the inflationary model is that during inflation some scalar
 field  is in a  ``wrong'' vacuum where the scale factor grows
 (quasi)exponentially with time.
The rapid expansion of the universe is almost  exponential, where
 sufficient time  forces the universe to be flat, and the correlated
 horizon to be large,
which is represented by an almost constant $H$ during inflation.\\
In this paper, $\rho +p $ is a non-negative constant of space-time.
If we write
\begin{equation}
H:=\sqrt{\frac{\rho + \Lambda}{3}}
,
\end{equation}
and solve (i) and (ii), we obtain 
\[
a(t)^2=\left(\frac{\cosh{H t}}{H}\right)^2\;\;\;\;\;\mbox{for } k=+1,
\]
\[
a(t)^2=\left(\mbox{constant}\cdot \exp{H t}\right)^2
\;\;\;\;\;\mbox{for } k=0,
\]
and
\begin{equation}
a(t)^2=\left(\frac{\sinh{H t}}{H}\right)^2\;\;\;\;\;\mbox{for } k=-1.
\end{equation}
One can find  that this space-time is four-dimensional maximally symmetric,
\begin{equation}
R_{\mu\nu\alpha\beta}
=H^2\left(g_{\mu\alpha}g_{\nu\beta}-g_{\nu\alpha}g_{\mu\beta}  \right) .
\end{equation}
To solve (iii) and (iv) we can find a solution for $\rho$ and  $p$
  (or $\phi$).\\
In the case of  arbitrarily $A$, $B$ and $\Lambda=-\frac{1}{2} c B$ with
 negative constant $c$ 
there is a solution that 
\[
\dot{\phi}=0\;.
\]
Then, $\rho=-p=\mbox{constant}=0$, which is static,
 and $H^2=\frac{\Lambda}{3}$. 
\section{One Loop Divergence of Effective Action}
\subsection{Back Ground Field method  and One-Loop Effective Action}
\hspace*{5mm}
We start with the back-ground field method \cite{Ab}. We split the fields
into background fields ($g_{\mu\nu}$, $\phi$)
and quantum fields ($h_{\mu\nu}$, $\varphi$): 
\begin{equation}
\phi \rightarrow \phi' = \phi+\varphi\;,\;\;\;\;\;\;\;\;\;\;\;\;\;\;
g_{\mu\nu} \rightarrow g'_{\mu\nu}= g_{\mu\nu}+ h_{\mu\nu}\;,        
\end{equation}
where the background fields ($g_{\mu\nu}$, $\phi$) are a solution of the
 classical equations of motion.
The one-loop effective action is given by the standard general expression,
\begin{equation}
\Gamma^{\mbox{\small 1-loop}}={i \over 2}\;\mbox{Tr} \ln {\hat{H}} 
- i\;\mbox{Tr}\ln {\hat{H}_{\mbox{gh}}},
\end{equation}
where $\hat{H}$ is the bilinear form of the action (\ref{action}) with 
an added gauge fixing term,  and $\hat{H}_{\mbox{gh}}$ is the bilinear
 form of
the ghosts action ($S_{\mbox{gh}}$).  To perform the calculations in the
 simplest
 way one needs to introduce a special form of the gauge
fixing term:
\begin{equation}
 S_{\mbox{gf}} = \int d^4 x
\sqrt{-g}\;\chi_{\mu}\;\frac{\alpha}{2}\;\chi^{\mu}, 
\end{equation}
where $\chi_{\mu} := \nabla_{\alpha} \bar{h}_{\mu}^{\,\alpha}+
\beta\nabla_{\mu}h+\gamma \nabla_{\mu} \varphi$, $h:=h_{\mu}^{\mu}$,
$\bar{h}_{\mu\nu}:=h_{\mu\nu}-\frac{1}{4}\;hg_{\mu\nu}$ and $\alpha,
\beta, \gamma$ are functions of the background dilaton, which can
be tuned for our purposes. For instance, if one choose these functions
as 
\begin{equation}
\alpha=-B\;\;,\;\;\;\;\beta=-\frac{1}{4}\;\;,
\;\;\;\;\gamma=-\frac{B_1}{B}\;,
\end{equation}
 the bilinear part of the action ($S+S_{\mbox{gf}}+ S_{\mbox{gh}}$)
and the operator ($\hat{H}$ and $\hat{H}_{\mbox{gh}}$) have an especially
simple (minimal) structure:
\begin{equation}
\left.\left(S + S_{\mbox{gf}} +
S_{\mbox{gh}}\right)\right|_{\mbox{bilinear}} =\int d^4 x \sqrt{-g}\;
\left({\Phi} \hat{H} {\Phi}^T + c_{\mu}\hat{H_{\mbox{gh}}c^{\mu}}
\right)\;,
\end{equation}
where
\begin{equation}
 \hat{H}=\hat{K}\Box +\hat{L}_{\rho}\nabla^{\rho}+ \hat{M}
\;\;\;\mbox{and}\;\;\;\;\ \hat{H}_{\mbox{gh}}= g^{\mu
\alpha}\Box+\gamma(\nabla^{\alpha}\phi)\nabla^{\mu} + \gamma
(\nabla^{\mu} \nabla^{\alpha} \phi) + R^{\mu \alpha} 
\end{equation}
Here, $\Phi=\left(\bar{h}_{\mu\nu},\;h,\; \varphi\right)$, $c_{\mu}$
stand for  ghosts and $T$ stands for transposition. The components of
 $\hat{H}$
have the following form:
\begin{equation}
\hat{K}=\left(
\begin{array}{ccc}
              \frac{B}{4} \delta^{\mu\nu ,\alpha \beta} & 0 & 0\\ 0 & -
              \frac{B}{16} & -\frac{B_1}{4} \\ 0 & -\frac{B_1}{4} &
              \frac{B_1^2}{2B} -A
\end{array}
\right)
\end{equation}
{\small
\[
\hat{L}^{\lambda}=
\]
\begin{equation}
\left(
\begin{array}{ccc}
              \frac{B_1}{4} \left(\delta^{\mu \nu \alpha \beta}
g^{\tau \lambda} + 2 g^{\nu \beta}\left(g^{\mu \tau } g^{\alpha
\lambda } - g^{\alpha \tau } g^{\mu \lambda }\right)\right) & -
\frac{B_1}{4} g^{\mu \tau} g^{\nu \lambda} & \left( \frac{B_2}{2}-A
\right) g^{\mu \tau} g^{\nu \lambda}\\ \frac{B_1}{4} g^{\alpha \tau}
g^{\beta \lambda} & -\frac{B_1}{16} g^{\tau \lambda} &
\left(\frac{A}{4} -\frac{5}{8} B_2 \right) g^{\tau \lambda}\\ \left( A
- \frac{B_2}{2}\right) g^{\alpha \tau} g^{\beta \lambda} & \left(
\frac{B_2}{8}-\frac{A}{4}\right) g^{\tau \lambda} &
\left(\frac{B_1^2}{2B} -A\right)_1 g^{\tau \lambda}
\end{array}
\right) (\nabla_{\tau}\phi)
\end{equation}
}
{\small
\[
\hat{M}=
\]
\begin{equation}
 \left(\!\!\!\!\!
\begin{array}{ccc}
    
    \begin{array}{l}
           
           \delta^{\mu \nu \alpha \beta} \left( \frac{B_1}{2}(\Box
                                         \phi) +\left( \frac{B_2}{2} -
                                         \frac{A}{4}\right)(\nabla
                                         \phi)^2 +\frac{B \Lambda}{2}
                                         \right) \\

         + g^{\nu \beta}\left( -B_1\left( \nabla^{\mu} \nabla^\alpha
                     \phi \right) +\left( A-B_2 \right)(\nabla^\mu
                     \phi)(\nabla^\alpha \phi) \right) \\

        +\frac{B}{4}\left( -\delta^{\mu \nu \alpha \beta} R + 2 g^{\nu
                    \beta} R^{\mu \alpha} +2R^{\mu \alpha \nu \beta}
                    \right)

   \end{array}

           \!\!\!\!\!\!\!\!\!\!\!\!\!\!
          & \!\!\!\!\!\!\!\!\!\!\!\!\!\! 0 \!\!\!\!\!\!\!\!\!\!\!\! 
          & \!\!\!\!\!\!\!\!\!\!\!\!

   \begin{array}{l}

           \frac{B_2}{2}\left( \nabla^{\mu} \nabla^{\nu} \phi \right)
        
             \\
      
           + \left( \frac{B_3}{2} - \frac{A_1}{2} \right) (\nabla^\mu
                       \phi)(\nabla^\nu \phi) \\

           - \frac{B_1}{2}R^{\mu \nu}

    \end{array}

 \\ \!\!\!\!\!\! & \!\!\!\!\!\!  \!\!\!\!\!\! & \!\!\!\!\!\!
          
 \\

              \frac{B_1}{4}\left( \nabla^{\alpha} \nabla^{\beta} \phi
            \right) + \frac{B_2}{4} (\nabla^\alpha \phi)(\nabla^\beta
            \phi)

           \!\!\!\!\!\!\!\!\!\!\!\! & \!\!\!\!\!\!\!\!\!\!\!\!
            -\frac{B \Lambda}{8} 
           \!\!\!\!\!\! & \!\!\!\!\!\!\!\!\!\!
            \begin{array}{l}

              -\frac{3}{8} B_2 (\Box \phi)
     
               \\
     
              + \left( \frac{A_1}{8} - \frac{3}{8}B_3 \right)(\nabla
\phi)^2
             
                \\
             
              + \frac{B_1}{8} R - \frac{(B \Lambda)_1}{2}
             
     \end{array}
 \\ 
 \!\!\!\!\!\! & \!\!\!\!\!\!\!\!\!\!\!\! &
 \!\!\!\!\!\! 
 \\
  
               A\left( \nabla^{\alpha} \nabla^{\beta} \phi \right) +
             \frac{A_1}{2} (\nabla^\alpha \phi)(\nabla^\beta \phi) -
             \frac{B_1}{2}R^{\alpha \beta}
          
            \!\! & \!\!
    
     \begin{array}{l}
     
             -\frac{A}{4} (\Box \phi)
    
                \\
        
            - \frac{A_1}{8} (\nabla \phi)^2
             
                 \\
           
            + \frac{B_1}{8} R - \frac{(B \Lambda)_1}{2}
           
     \end{array}
  
              \!\!\!\!\!\!&\!\!\!\!\!\!\!\!
    
      \begin{array}{l}
    
             -A_1(\Box \phi)
   
                  \\
       
            -\frac{A_2}{2}(\nabla \phi)^2
      
                   \\
           
            + \frac{B_2}{2} R - (B \Lambda)_2
         
      \end{array}
\end{array}
\!\!\!\!\!\! \right)
\end{equation}
}
\vspace*{2mm}\\
The next problem is to separate the divergent part of $\mbox{Tr}\ln\hat{H}$.
To do this we rewrite this trace as
\begin{equation} \mbox{Tr} \ln\hat{H} =\mbox{Tr}\ln\hat{K}
+ \mbox{Tr}\ln\left(\hat{1}\Box +
\hat{K}^{-1} \hat{L}^{\mu}\nabla_\mu +\hat{K}^{-1}\hat{M} \right)\;.
\label{trlnH}
\end{equation}
We note that the first term does not contribute to the
divergences. 
\subsection{One Loop Divergence  by Schwinger-DeWitt technique }
\hspace*{5mm}
We now  explore the second term in eq.(\ref{trlnH}) 
which has the standard minimal form, and can be easily estimated 
by using the standard Schwinger-DeWitt method \cite{De,BOS}. 
The general off shell structure of a one-loop divergence of the effective
 action
is as  follows \cite{ST1}:
\[
\Gamma_{\mbox{\small div}}^{\mbox{\small 1-loop}}=\frac{1}{16 \pi^2
\epsilon} \int d^4x\sqrt{-g} \left[ c_e G + c_c
C_{\mu\nu\alpha\beta}C^{\mu\nu\alpha\beta} +c_r R^2 +c_7 R +c_{12} \right.
\]
\[
 + c_4 R(\nabla \phi)^2 + c_5 R(\Box \phi )+ c_6 R^{\mu
\nu}(\nabla_{\mu} \phi)(\nabla_\nu \phi) + c_{11} (\nabla
\phi)^2  
\] 
\begin{equation}
\left. 
+ c_8 (\nabla \phi)^4 + c_9
(\nabla \phi)^2(\Box \phi)+ c_{10} (\Box \phi)^2  +(\mbox{surface term})
\right]
 \;,
\end{equation}
where $\epsilon:=D-4$ is a dimensional parameter; $c_e, c_c, c_r, $
and $c_4,\cdots c_{12}$ are some functions of $A,B,\Lambda$.\\
$G=R_{\mu\nu\alpha\beta}R^{\mu\nu\alpha\beta}-4R_{\mu\nu}R^{\mu\nu}+R^2$
is the Gauss-Bonnet topological invariant 
and $C_{\mu\nu\alpha\beta}$ is the Weyl tenser 
$\left(C_{\mu\nu\alpha\beta}C^{\mu\nu\alpha\beta}
=R_{\mu\nu\alpha\beta}R^{\mu\nu\alpha\beta}-2R_{\mu\nu}R^{\mu\nu}
+\frac{1}{3}R^2\right)$.\\
In a homogeneous and isotropic space-time this structure can be written
as follows.
\[
\Gamma_{\mbox{\small div, R-W}}^{\mbox{\small 1-loop}}
\]
\[
:=\frac{1}{16
\pi^2 \epsilon} \int d^3x dt\sqrt{-g} \left[ a_1 \left(\kappa_s
+\kappa_t \right)^2 + a_2 \kappa_s \kappa_t
+\frac{\Lambda}{3}a_3\left(\kappa_s +\kappa_t \right) +
\left(\frac{\Lambda}{3}\right)^2 a_4 \right.
\] 
\begin{equation}
\left.  + \left(b_1 \left(\kappa_s + \kappa_t \right) +b_2
\kappa_s + \frac{\Lambda}{3}b_3\right)\dot{\phi}^2+
\frac{\Lambda}{3}b_4 \ddot{\phi}+ c_1 \dot{\phi}^4 +c_2 \dot{\phi}^2
\ddot{\phi} +c_3 \ddot{\phi}^2 \right] +(\mbox{s.t.})\;,  
\label{rw}
\end{equation}
where $\kappa_s:=\frac{\ddot{a}}{a}$, $\kappa_t
:=\left(\frac{\dot{a}}{a}\right)^2
+\frac{k}{a^2} $, and the explicit structure of $a_1 \cdots c_3$ is
expressed in Appendix.
\subsection{Divergence on Inflationary Back Ground 
 and Renormalization} 
\hspace*{5mm}
Here, we consider an inflationary universe where $\kappa_s
    =\kappa_t=H^2=\frac{\Lambda}{3}=-\frac{c B}{6}$.
Thus the structure of the divergence terms is very simple, such that
\[
\Gamma_{\mbox{\small div, D-S}}^{\mbox{1-loop}}:= \Gamma_{\mbox{\small
div, R-W}}^{\mbox{1-loop}}\left|_{R_{\mu\nu\alpha\beta}
=\frac{\Lambda}{3}\left(g_{\mu\alpha}g_{\nu\beta}
-g_{\nu\alpha}g_{\mu\beta}\right),\;\dot{\phi}=0}
\right.
\]
\begin{equation}
=\frac{1}{16 \pi^2 \epsilon}\int d^4x\sqrt{-g}\left[ -
\frac{371}{90}\Lambda(\phi)^2 \right]\;.
\end{equation}
Remarkably,  all of the dependences of $A(\phi)$ and the derivative of
 $B(\phi)$ 
canceled out (or all $\delta$, $\omega$ dependence was cancelled). 
The on-shell structure of the effective action is a very simple, such that
  we can easily renormalize that divergence by
only renormalizing the function $B(\phi)$ (or $\Lambda(\phi)$);  $c$
and $\phi$ do not have to be renormalized.
\begin{equation}
B_0(\phi):=\mu^{ \frac{\epsilon}{2}} \left(1-\frac{c}{16 \pi^2
\epsilon}\frac{371}{180}\right)B(\phi)\;,
\end{equation}
where $\mu$ is the renormalization scale and $B_0$ is a bare function.
\section{Conclusion and Discussion}
\hspace*{5mm}
In this  paper we considered the general action (\ref{action}) on
 homogeneous and isotropic four-dimensional background space-time, and used
 the Robertson-Walker metric, which is a realistic model of a hot universe.
It is a spatially  maximally symmetric space-time in which all quantities
 depend only on time.
First, we considered this action at the  classical level.
We found that the classical equations of motion concerning the  metric are
 the ``Hubble'' equation and the energy momentum conservation law, and that
 the equation of motion of dilaton is the ``state'' equation between the
  energy density and a pressure driven by dilaton. Since we have four
 parameters and five equations, it seems at first that there are no
 solutions. 
However, if we set $\Lambda=-\frac{1}{2}c B $ with a negative constant
 ($c$) on the inflationary background, we were able to find a solution with
 constant dilaton, zero density, zero pressure,
 and $H^2=\frac{\Lambda}{3}$. This solution is independent of time, and the
 background is the maximally symmetric four-dimensional space-time.
Next, we considered this action at the  quantum level.
A one-loop calculation was carried out for the model (\ref{action}) using
 the background field method. This calculation is formally the same as
 \cite{ST1} but background fields must follow a solution of the equations
 of motion on homogeneous and isotropic space-time. We pulled a bilinear
 form out of action (\ref{action}) with a gauge fixing term added, and out
 of the ghost action. Such a form is sufficient to calculate the effective
 action at one-loop levels. We have fixed a gauge for the minimal one to
 cancel the derivative terms, except for the d'Alembertian terms; we were
 then able to apply the standard Schwinger-DeWitt method to estimate the
 divergence of effective action. We found that the divergence is
 constructed by the scale factor of universe, the dilaton field,
 $\delta(=-\frac{B_1}{B})$,
 $\omega(=\frac{1}{B} \left(A-\frac{3}{2}\frac{B_1^2}{B} \right))$, and
 their time derivatives. The spatial distinction of the universe  between
 open, flat and closed depends only on $\kappa_t$.
In this background the structure of the divergence is so complicated that
 it seems that the divergence of the effective action cannot be removed by
 renormalizing any quantity in the theory. 
If we consider an inflationary background (maximal symmetric one )
 especially, the structure of the divergence, however, turned out  to be
 miraculously simple, such that there is only one term, which depends only
 on the cosmological function $(\Lambda(\phi)$) under a general function
 form of $A(\phi),B(\phi)$; also, the divergence of the effective action
 can be removed by multiplicatively renormalizing only  $\Lambda(\phi)$.
 All of the dependences of $A(\phi)$ and the derivative of $B(\phi)$ were
 canceled. 
Such miraculous cancellations cannot be accidental.
This suggests that the cancellations also occur  in the higher loop
 structure of the divergence of the effective action during inflation, and
 that these cancellations  may represent an effective appearance of some
 symmetry of the un-constructed complete quantum gravity on the
 inflationary background. 
Considering the renormalizable condition in the case of  the present paper,
 in the case of \cite{ST1} and in the case of pure gravity,  we can find
 that all cases are the same; namely, in all cases the one-loop effective
 action is finite, or renormalizable, in the case of the absence or
 presence of  a cosmological term, respectively. 
However, there is a difference: in the case of \cite{ST1} we must tune the
 functions $ A,B,\Lambda$, and arbitrary backgrounds are allowed. On the
 other hand, in the case of the present paper, arbitrary functions
 ($ A,B,\Lambda$) are allowed, and backgrounds are restricted to a
 maximally symmetric one. 
It seems that the property between these models gives a cosmological
 vision between before inflation, during inflation and after inflation,
 namely an energy region where the dilaton has an important role with
 special coupling to the metric in the case of \cite{ST1}, the region where
 the dilaton is constant with general coupling to the metric in the case of
 the present paper, and where the dilaton is decoupled to the metric in the
 case of pure gravity. However,in order to reliably discuss the transitions
 between these phases  we must find a complete quantum gravity, like a
 string theory, which can explain these phases in a unified way.
\section*{Acknowledgements}
\hspace*{5mm}
The author is grateful to H.Kawai and the entire Department of Theoretical
 and Computational Physics at KEK for the stimulating discussions.
He is also grateful to I.L.Shapiro for various suggestions by e-mail.
He is thankful to T.Muta and the entire Department of Particle Physics at
 Hiroshima University for the interesting discussions, especially regarding
 their work \cite{IMM}.
\appendix
\section{Appendix}
The coefficients of the structure of divergence in equation (\ref{rw}) are
 the following( with notation $\delta(\phi):=-\frac{B_1}{B}$ and
$\omega(\phi):=\frac{1}{B}\left(A-\frac{3}{2}\frac{B_1^2}{B}\right)$):
{\small
\[
a_1= {\displaystyle \frac {19}{2}} + 2\,{\displaystyle \frac {1}{{
\omega}^{2}}} + 6\,{\displaystyle \frac {1}{{ \omega}}} + 4\,
{\displaystyle \frac {{{ \delta}_{1}}}{{ \delta}^{2}\,{ \omega}^{ 2}}}
+ 2\,{\displaystyle \frac {{{ \delta}_{1}}}{{ \delta}^{2}\,{ \omega}}}
+ 2\,{\displaystyle \frac {{{ \delta}_{1}}^{2}}{{ \delta}^{4}\,{
\omega}^{2}}} \;,
\]
\[
a_2= {\displaystyle \frac {99}{5}} +
8\,{\displaystyle \frac {1}{{ \omega}}}\;,
\]
\[
a_3= 26 + {\displaystyle \frac {4}{3}}\,{\displaystyle \frac {1}{{
\omega}^{2}}} + {\displaystyle \frac {14}{3}}\,{\displaystyle \frac
{1}{{ \omega}}} + {\displaystyle \frac {8}{3}}\, {\displaystyle \frac
{{{ \delta}_{1}}}{{ \delta}^{2}\,{ \omega}^{ 2}}} + {\displaystyle
\frac {2}{3}}\,{\displaystyle \frac {{{ \delta}_{1}}}{{ \delta}^{2}\,
{ \omega}}} + {\displaystyle \frac { 4}{3}}\,{\displaystyle \frac {{{
\delta}_{1}}^{2}}{{ \delta}^{4} \,{ \omega}^{2}}} \;,
\]
\[ 
a_4= 
5 +{\displaystyle \frac {2}{9}}\,{\displaystyle \frac {1}{{ \omega}^{2}}}
+ {\displaystyle \frac {2}{3}}\,{\displaystyle \frac {1}{{ \omega}}} +
{\displaystyle \frac {4}{9}}\, {\displaystyle \frac {{{
\delta}_{1}}}{{ \delta}^{2}\,{ \omega}^{ 2}}} + {\displaystyle \frac
{2}{9}}\,{\displaystyle \frac {{{ \delta}_{1}}^{2}}{{ \delta}^{4}\,{
\omega}^{2}}}\;,
\]
\begin{eqnarray*}
b_1= \lefteqn{ - \,{\displaystyle \frac {155}{4}}\,{ \delta}^{2} -
{\displaystyle \frac {{ \delta}^{2}}{{ \omega}^{2}}} - {\displaystyle
\frac {7}{2}}\,{\displaystyle \frac {{ \delta}^{2} }{{ \omega}}} +
6\,{ \delta}^{2}\,{ \omega} + 18\,{{ \delta}_{1}} + 9\,{\displaystyle
\frac {{{ \delta}_{1}}}{{ \omega}}} + {\displaystyle \frac {{{
\delta}_{1}}^{2}}{{ \delta}^{2}}} + 3\, {\displaystyle \frac {{{
\delta}_{1}}^{2}}{{ \delta}^{2}\,{ \omega}^{2}}} + 2\,{\displaystyle
\frac {{{ \delta}_{1}}^{3}}{{ \delta}^{4}\,{ \omega}^{2}}} +
2\,{\displaystyle \frac {{{ \delta }_{1}}^{3}}{{ \delta}^{4}\,{
\omega}}}} \\ & & \mbox{} + 3\,{\displaystyle \frac {{ \delta}\,{{
\omega}_{1} }}{{ \omega}^{2}}} + 4\,{\displaystyle \frac {{
\delta}\,{{ \omega}_{1}}}{{ \omega}}} + {\displaystyle \frac {{{
\delta}_{1}} \,{{ \omega}_{1}}}{{ \delta}\,{ \omega}}} +
2\,{\displaystyle \frac {{{ \delta}_{1}}^{2}\,{{ \omega}_{1}}}{{
\delta}^{3}\,{ \omega}^{2}}} + {\displaystyle \frac
{1}{2}}\,{\displaystyle \frac {{{ \omega}_{1}}^{2}}{{ \omega}^{3}}} +
{\displaystyle \frac {1}{4}}\,{\displaystyle \frac {{{
\omega}_{1}}^{2}}{{ \omega}^{2}}} + {\displaystyle \frac
{1}{2}}\,{\displaystyle \frac {{{ \delta}_{1}}\,{{ \omega}_{1}}^{2}}
{{ \delta}^{2}\,{ \omega}^{3}}}\;,\mbox{\hspace{33pt}}
\end{eqnarray*}
\[
b_2=
3\,{ \delta}^{2} - 3\,{\displaystyle \frac {{ \delta}^{2}}{{ 
\omega}}} + 6\,{{ \delta}_{1}} + 7\,{\displaystyle \frac {{{ 
\delta}_{1}}}{{ \omega}}} + 4\,{\displaystyle \frac {{ \delta}\,{
{ \omega}_{1}}}{{ \omega}^{2}}} + 6\,{\displaystyle \frac {{ 
\delta}\,{{ \omega}_{1}}}{{ \omega}}} - {\displaystyle \frac {{{ 
\delta}_{1}}\,{{ \omega}_{1}}}{{ \delta}\,{ \omega}^{2}}}\;,
\]
\begin{eqnarray*}
b_3=
\lefteqn{{\displaystyle \frac {4}{3}}\,{\displaystyle \frac {{ 
\delta}\,{{ \omega}_{1}}}{{ \omega}^{3}}} - {\displaystyle 
\frac {7}{2}}\,{ \delta}^{2} + {\displaystyle \frac {1}{3}}\,
{\displaystyle \frac {{ \delta}^{2}}{{ \omega}^{2}}} + 
{\displaystyle \frac {3}{2}}\,{\displaystyle \frac {{ \delta}^{2}
}{{ \omega}}} + 3\,{ \delta}^{2}\,{ \omega} + {\displaystyle 
\frac {3}{2}}\,{\displaystyle \frac {{{ \delta}_{1}}\,{{ \omega}
_{1}}}{{ \delta}\,{ \omega}^{2}}} + 2\,{\displaystyle \frac {{{ 
\delta}_{1}}^{2}\,{{ \omega}_{1}}}{{ \delta}^{3}\,{ \omega}^{2}}}
 + {\displaystyle \frac {5}{6}}\,{\displaystyle \frac {{{ \delta}
_{1}}\,{{ \omega}_{1}}^{2}}{{ \delta}^{2}\,{ \omega}^{3}}} - 
{\displaystyle \frac {4}{3}}\,{\displaystyle \frac {{{ \delta}_{1
}}\,{{ \delta}_{2}}}{{ \delta}^{3}\,{ \omega}^{2}}}} \\
 & & \mbox{} - {\displaystyle \frac {4}{3}}\,{\displaystyle 
\frac {{{ \delta}_{1}}\,{{ \delta}_{2}}}{{ \delta}^{3}\,{ \omega}
}} - {\displaystyle \frac {1}{3}}\,{\displaystyle \frac {{{ 
\omega}_{1}}\,{{ \delta}_{2}}}{{ \delta}^{2}\,{ \omega}^{2}}} + 
{\displaystyle \frac {4}{3}}\,{\displaystyle \frac {{{ \delta}_{1
}}^{2}\,{{ \omega}_{1}}}{{ \delta}^{3}\,{ \omega}^{3}}} + 
{\displaystyle \frac {8}{3}}\,{\displaystyle \frac {{{ \delta}_{1
}}\,{{ \omega}_{1}}}{{ \delta}\,{ \omega}^{3}}} - {\displaystyle 
\frac {1}{3}}\,{\displaystyle \frac {{{ \delta}_{1}}\,{{ \omega}
_{2}}}{{ \delta}^{2}\,{ \omega}^{2}}} + 3\,{\displaystyle \frac {
{{ \delta}_{1}}^{2}}{{ \delta}^{2}\,{ \omega}^{2}}} + 
{\displaystyle \frac {8}{3}}\,{\displaystyle \frac {{{ \delta}_{1
}}^{3}}{{ \delta}^{4}\,{ \omega}^{2}}} + {\displaystyle \frac {8
}{3}}\,{\displaystyle \frac {{{ \delta}_{1}}^{3}}{{ \delta}^{4}\,
{ \omega}}} \\
 & & \mbox{} + {\displaystyle \frac {7}{3}}\,{\displaystyle 
\frac {{ \delta}\,{{ \omega}_{1}}}{{ \omega}^{2}}} + 
{\displaystyle \frac {1}{2}}\,{\displaystyle \frac {{ \delta}\,{{
 \omega}_{1}}}{{ \omega}}} - {\displaystyle \frac {4}{3}}\,
{\displaystyle \frac {{{ \delta}_{2}}}{{ \delta}\,{ \omega}^{2}}}
 - {\displaystyle \frac {{{ \delta}_{2}}}{{ \delta}\,{ \omega}}}
 + {\displaystyle \frac {4}{3}}\,{\displaystyle \frac {{{ \delta}
_{1}}^{2}}{{ \delta}^{2}\,{ \omega}}} + {\displaystyle \frac {{{ 
\delta}_{1}}}{{ \omega}}} + {\displaystyle \frac {5}{6}}\,
{\displaystyle \frac {{{ \omega}_{1}}^{2}}{{ \omega}^{3}}} + 
{\displaystyle \frac {2}{3}}\,{\displaystyle \frac {{{ \delta}_{1
}}}{{ \omega}^{2}}} - {\displaystyle \frac {1}{3}}\,
{\displaystyle \frac {{{ \omega}_{2}}}{{ \omega}^{2}}}\;,
\end{eqnarray*}
\[
b_4=
{\displaystyle \frac {31}{2}}\,{ \delta} + 2\,{\displaystyle 
\frac {{ \delta}}{{ \omega}^{2}}} + 6\,{\displaystyle \frac {{ 
\delta}}{{ \omega}}} + {\displaystyle \frac {{{ \delta}_{1}}}{{ 
\delta}}} + 4\,{\displaystyle \frac {{{ \delta}_{1}}}{{ \delta}\,
{ \omega}^{2}}} + 4\,{\displaystyle \frac {{{ \delta}_{1}}}{{ 
\delta}\,{ \omega}}} + 2\,{\displaystyle \frac {{{ \delta}_{1}}^{
2}}{{ \delta}^{3}\,{ \omega}^{2}}} + 2\,{\displaystyle \frac {{{ 
\delta}_{1}}^{2}}{{ \delta}^{3}\,{ \omega}}} + {\displaystyle 
\frac {{{ \omega}_{1}}}{{ \omega}^{2}}} + {\displaystyle \frac {1
}{2}}\,{\displaystyle \frac {{{ \omega}_{1}}}{{ \omega}}} + 
{\displaystyle \frac {{{ \delta}_{1}}\,{{ \omega}_{1}}}{{ \delta}
^{2}\,{ \omega}^{2}}}\;,
\]
\begin{eqnarray*}
c_1=
\lefteqn{{\displaystyle \frac {33}{16}}\,{ \delta}^{4}\,{ \omega}
 + {\displaystyle \frac {201}{32}}\,{ \delta}^{4} + 
{\displaystyle \frac {45}{4}}\,{ \delta}^{4}\,{ \omega}^{2} + 
{\displaystyle \frac {1}{8}}\,{\displaystyle \frac {{ \delta}^{4}
}{{ \omega}^{2}}} + {\displaystyle \frac {1}{2}}\,{\displaystyle 
\frac {{ \delta}^{4}}{{ \omega}}} - {\displaystyle \frac {1}{6}}
\,{\displaystyle \frac {{{ \delta}_{1}}\,{{ \omega}_{2}}}{{ 
\omega}^{2}}} - {\displaystyle \frac {1}{6}}\,{\displaystyle 
\frac {{{ \omega}_{1}}\,{{ \delta}_{2}}}{{ \omega}^{2}}} + 
{\displaystyle \frac {3}{4}}\,{\displaystyle \frac {{{ \delta}_{1
}}\,{{ \delta}_{2}}}{{ \delta}}} + {\displaystyle \frac {3}{2}}\,
{ \delta}\,{ \omega}\,{{ \delta}_{2}}} \\
 & & \mbox{} + {\displaystyle \frac {25}{24}}\,{\displaystyle 
\frac {{ \delta}\,{{ \delta}_{2}}}{{ \omega}}} - {\displaystyle 
\frac {19}{16}}\,{\displaystyle \frac {{{ \delta}_{1}}\,{{ \omega
}_{1}}^{2}}{{ \omega}^{2}}} + {\displaystyle \frac {11}{24}}\,
{\displaystyle \frac {{{ \delta}_{1}}\,{{ \omega}_{1}}^{2}}{{ 
\omega}^{3}}} - {\displaystyle \frac {13}{16}}\,{\displaystyle 
\frac {{ \delta}^{2}\,{{ \omega}_{1}}^{2}}{{ \omega}^{2}}} - 
{\displaystyle \frac {19}{24}}\,{\displaystyle \frac {{ \delta}^{
2}\,{{ \omega}_{1}}^{2}}{{ \omega}^{3}}} - {\displaystyle \frac {
11}{16}}\,{\displaystyle \frac {{ \delta}^{3}\,{{ \omega}_{1}}}{{
 \omega}}} + {\displaystyle \frac {5}{12}}\,{\displaystyle 
\frac {{ \delta}^{3}\,{{ \omega}_{1}}}{{ \omega}^{2}}} \\
 & & \mbox{} + {\displaystyle \frac {{{ \delta}_{1}}^{4}}{{ 
\delta}^{4}\,{ \omega}}} + {\displaystyle \frac {1}{2}}\,
{\displaystyle \frac {{{ \delta}_{1}}^{4}}{{ \delta}^{4}\,{ 
\omega}^{2}}} - {\displaystyle \frac {5}{4}}\,{\displaystyle 
\frac {{{ \delta}_{1}}^{3}}{{ \delta}^{2}\,{ \omega}}} + 
{\displaystyle \frac {{ \delta}^{2}\,{{ \omega}_{2}}}{{ \omega}}}
 - {\displaystyle \frac {9}{4}}\,{ \delta}\,{{ \delta}_{1}}\,{{ 
\omega}_{1}} - {\displaystyle \frac {27}{4}}\,{ \delta}^{2}\,{ 
\omega}\,{{ \delta}_{1}} - {\displaystyle \frac {1}{4}}\,
{\displaystyle \frac {{ \delta}^{2}\,{{ \delta}_{1}}}{{ \omega}^{
2}}} + {\displaystyle \frac {1}{3}}\,{\displaystyle \frac {{ 
\delta}^{2}\,{{ \omega}_{2}}}{{ \omega}^{2}}} \\
 & & \mbox{} + {\displaystyle \frac {5}{8}}\,{\displaystyle 
\frac {{{ \omega}_{1}}\,{{ \delta}_{2}}}{{ \omega}}} - 
{\displaystyle \frac {21}{16}}\,{\displaystyle \frac {{ \delta}^{
2}\,{{ \omega}_{1}}^{2}}{{ \omega}}} + {\displaystyle \frac {1}{2
}}\,{\displaystyle \frac {{{ \delta}_{1}}^{3}}{{ \delta}^{2}\,{ 
\omega}^{2}}} + {\displaystyle \frac {3}{4}}\,{ \delta}^{2}\,{{ 
\omega}_{2}} - {\displaystyle \frac {13}{4}}\,{\displaystyle 
\frac {{{ \delta}_{1}}^{3}}{{ \delta}^{2}}} + {\displaystyle 
\frac {1}{2}}\,{\displaystyle \frac {{{ \delta}_{1}}^{4}}{{ 
\delta}^{4}}} + {\displaystyle \frac {1}{32}}\,{\displaystyle 
\frac {{{ \omega}_{1}}^{4}}{{ \omega}^{4}}} + {\displaystyle 
\frac {11}{8}}\,{ \delta}\,{{ \delta}_{2}} \\
 & & \mbox{} - {\displaystyle \frac {15}{4}}\,{ \omega}\,{{ 
\delta}_{1}}^{2} - 6\,{ \delta}^{3}\,{{ \omega}_{1}} - 
{\displaystyle \frac {3}{8}}\,{\displaystyle \frac {{{ \delta}_{1
}}^{2}}{{ \omega}^{2}}} + {\displaystyle \frac {25}{12}}\,
{\displaystyle \frac {{{ \delta}_{1}}^{2}}{{ \omega}}} - 
{\displaystyle \frac {367}{16}}\,{ \delta}^{2}\,{{ \delta}_{1}}
 + 10\,{{ \delta}_{1}}^{2} - {\displaystyle \frac {1}{8}}\,
{\displaystyle \frac {{ \delta}\,{{ \omega}_{1}}^{3}}{{ \omega}^{
3}}} - {\displaystyle \frac {1}{4}}\,{\displaystyle \frac {{{ 
\delta}_{1}}\,{{ \omega}_{2}}}{{ \omega}}} \\
 & & \mbox{} - {\displaystyle \frac {45}{16}}\,{\displaystyle 
\frac {{ \delta}^{2}\,{{ \delta}_{1}}}{{ \omega}}} - 
{\displaystyle \frac {17}{16}}\,{\displaystyle \frac {{ \delta}\,
{{ \delta}_{1}}\,{{ \omega}_{1}}}{{ \omega}^{2}}} + 
{\displaystyle \frac {15}{4}}\,{\displaystyle \frac {{ \delta}\,{
{ \delta}_{1}}\,{{ \omega}_{1}}}{{ \omega}}} + {\displaystyle 
\frac {1}{8}}\,{\displaystyle \frac {{{ \delta}_{1}}^{2}\,{{ 
\omega}_{1}}}{{ \delta}\,{ \omega}^{2}}} - {\displaystyle \frac {
17}{4}}\,{\displaystyle \frac {{{ \delta}_{1}}^{2}\,{{ \omega}_{1
}}}{{ \delta}\,{ \omega}}} - {\displaystyle \frac {1}{4}}\,
{\displaystyle \frac {{{ \delta}_{1}}\,{{ \delta}_{2}}}{{ \delta}
\,{ \omega}}} + {\displaystyle \frac {1}{4}}\,{\displaystyle 
\frac {{{ \delta}_{1}}\,{{ \omega}_{1}}^{3}}{{ \delta}\,{ \omega}
^{3}}} \\
 & & \mbox{} + {\displaystyle \frac {3}{4}}\,{\displaystyle 
\frac {{{ \delta}_{1}}^{2}\,{{ \omega}_{1}}^{2}}{{ \delta}^{2}\,{
 \omega}^{2}}} + {\displaystyle \frac {1}{4}}\,{\displaystyle 
\frac {{{ \delta}_{1}}^{2}\,{{ \omega}_{1}}^{2}}{{ \delta}^{2}\,{
 \omega}^{3}}} + {\displaystyle \frac {{{ \delta}_{1}}^{3}\,{{ 
\omega}_{1}}}{{ \delta}^{3}\,{ \omega}}} + {\displaystyle \frac {
{{ \delta}_{1}}^{3}\,{{ \omega}_{1}}}{{ \delta}^{3}\,{ \omega}^{2
}}}\;,
\end{eqnarray*}
\begin{eqnarray*}
c_2=
\lefteqn{ - \,{\displaystyle \frac {105}{8}}\,{ \delta}^{3}\,{ 
\omega} - {\displaystyle \frac {47}{2}}\,{ \delta}^{3} + 
{\displaystyle \frac {{{ \delta}_{1}}^{3}}{{ \delta}^{3}\,{ 
\omega}^{2}}} + {\displaystyle \frac {35}{8}}\,{\displaystyle 
\frac {{ \delta}\,{{ \delta}_{1}}}{{ \omega}}} - {\displaystyle 
\frac {1}{2}}\,{\displaystyle \frac {{ \delta}^{3}}{{ \omega}^{2}
}} + {\displaystyle \frac {1}{4}}\,{\displaystyle \frac {{{ 
\delta}_{1}}\,{{ \omega}_{1}}}{{ \omega}^{2}}} - {\displaystyle 
\frac {11}{8}}\,{\displaystyle \frac {{{ \delta}_{1}}\,{{ \omega}
_{1}}}{{ \omega}}} - {\displaystyle \frac {7}{4}}\,
{\displaystyle \frac {{ \delta}^{3}}{{ \omega}}} - 
{\displaystyle \frac {3}{4}}\,{ \delta}\,{ \omega}\,{{ \delta}_{1
}}} \\
 & & \mbox{} + {\displaystyle \frac {3}{2}}\,{\displaystyle 
\frac {{{ \delta}_{1}}^{2}}{{ \delta}\,{ \omega}^{2}}} - 
{\displaystyle \frac {1}{4}}\,{\displaystyle \frac {{{ \delta}_{1
}}^{2}}{{ \delta}\,{ \omega}}} + 2\,{\displaystyle \frac {{{ 
\delta}_{1}}^{3}}{{ \delta}^{3}\,{ \omega}}} + {\displaystyle 
\frac {1}{4}}\,{\displaystyle \frac {{ \delta}^{2}\,{{ \omega}_{1
}}}{{ \omega}^{2}}} + {\displaystyle \frac {17}{8}}\,
{\displaystyle \frac {{ \delta}^{2}\,{{ \omega}_{1}}}{{ \omega}}}
 + {\displaystyle \frac {1}{4}}\,{\displaystyle \frac {{ \delta}
\,{{ \omega}_{1}}^{2}}{{ \omega}^{3}}} - {\displaystyle \frac {1
}{8}}\,{\displaystyle \frac {{ \delta}\,{{ \omega}_{1}}^{2}}{{ 
\omega}^{2}}} + {\displaystyle \frac {1}{8}}\,{\displaystyle 
\frac {{{ \omega}_{1}}^{3}}{{ \omega}^{3}}} \\
 & & \mbox{} - {\displaystyle \frac {3}{8}}\,{ \delta}^{2}\,{{ 
\omega}_{1}} + {\displaystyle \frac {131}{8}}\,{ \delta}\,{{ 
\delta}_{1}} - {\displaystyle \frac {9}{4}}\,{\displaystyle 
\frac {{{ \delta}_{1}}^{2}}{{ \delta}}} + {\displaystyle \frac {{
{ \delta}_{1}}^{3}}{{ \delta}^{3}}} + {{ \delta}_{2}} + 
{\displaystyle \frac {3}{2}}\,{\displaystyle \frac {{{ \delta}_{1
}}^{2}\,{{ \omega}_{1}}}{{ \delta}^{2}\,{ \omega}}} + 
{\displaystyle \frac {1}{4}}\,{\displaystyle \frac {{{ \delta}_{1
}}\,{{ \omega}_{1}}^{2}}{{ \delta}\,{ \omega}^{3}}} + 
{\displaystyle \frac {3}{2}}\,{\displaystyle \frac {{{ \delta}_{1
}}^{2}\,{{ \omega}_{1}}}{{ \delta}^{2}\,{ \omega}^{2}}} \\
 & & \mbox{} + {\displaystyle \frac {3}{4}}\,{\displaystyle 
\frac {{{ \delta}_{1}}\,{{ \omega}_{1}}^{2}}{{ \delta}\,{ \omega}
^{2}}}\;,
\end{eqnarray*}
\begin{eqnarray*}
c_3=
\lefteqn{{\displaystyle \frac {43}{8}}\,{ \delta}^{2} + 
{\displaystyle \frac {1}{2}}\,{\displaystyle \frac {{ \delta}^{2}
}{{ \omega}^{2}}} + {\displaystyle \frac {3}{2}}\,{\displaystyle 
\frac {{ \delta}^{2}}{{ \omega}}} + {\displaystyle \frac {3}{2}}
\,{{ \delta}_{1}} + {\displaystyle \frac {{{ \delta}_{1}}}{{ 
\omega}^{2}}} + {\displaystyle \frac {3}{2}}\,{\displaystyle 
\frac {{{ \delta}_{1}}}{{ \omega}}} + {\displaystyle \frac {1}{2
}}\,{\displaystyle \frac {{{ \delta}_{1}}^{2}}{{ \delta}^{2}}} + 
{\displaystyle \frac {1}{2}}\,{\displaystyle \frac {{{ \delta}_{1
}}^{2}}{{ \delta}^{2}\,{ \omega}^{2}}} + {\displaystyle \frac {{{
 \delta}_{1}}^{2}}{{ \delta}^{2}\,{ \omega}}} + {\displaystyle 
\frac {1}{2}}\,{\displaystyle \frac {{ \delta}\,{{ \omega}_{1}}}{
{ \omega}^{2}}} + {\displaystyle \frac {1}{4}}\,{\displaystyle 
\frac {{ \delta}\,{{ \omega}_{1}}}{{ \omega}}}} \\
 & & \mbox{} + {\displaystyle \frac {1}{2}}\,{\displaystyle 
\frac {{{ \delta}_{1}}\,{{ \omega}_{1}}}{{ \delta}\,{ \omega}^{2}
}} + {\displaystyle \frac {1}{2}}\,{\displaystyle \frac {{{ 
\delta}_{1}}\,{{ \omega}_{1}}}{{ \delta}\,{ \omega}}} + 
{\displaystyle \frac {1}{8}}\,{\displaystyle \frac {{{ \omega}_{1
}}^{2}}{{ \omega}^{2}}}\mbox{\hspace{237pt}}\;.
\end{eqnarray*}
}


\begin{thebibliography}{99}
\bibitem{GSW} M. B. Green, J. H. Schwarz and E. Witten, 
{\sl Superstring Theory}, 
(Cambridge University Press, Cambridge, 1987). 

\bibitem{St} K. S. Stelle, 
{\sl Renormalization of the Higher Derivative Quantum Gravity}, 
Phys. Rev. {\bf 16D}, (1977), 953. 

\bibitem{BOS} I. L. Buchbinder, S. D. Odintsov and I. L. Shapiro, 
{\sl Effective Action in Quantum Gravity}, 
(IOP, Bristol, 1992).


\bibitem{We} S. Weinberg, 
{\sl Gravitation and Cosmology}, 
(J. Wiley and Sons, 1972). 


\bibitem{Gu} A. H. Guth, 
{\sl A Possible Solution to the Horizon and Flatness Problems}, 
Phys. Rev. {\bf D23}, (1981), 347.  

\bibitem{Li1} A. Linde, 
{\sl A New Inflationary Universe Scenario: A Possible Solution of the
 Horizon, Flatness, Homogeneity, Isotropy and Primordial Monopole Problems}
, 
Phys. Lett. {\bf 108B},(1982),389.   

\bibitem{Li2} A. Linde, 
{\sl Chaotic Inflation}, 
Phys. Lett. {\bf 129B}, (1983), 177. 

\bibitem{ST} P. J. Steinhardt and M. Turner,
{\sl  A Prescription for Successful New  Inflation}, 
Phys. Rev. {\bf D29}, (1984), 2162.


\bibitem{Li3} A. Linde, 
{\sl Particle Physics and Inflationary Cosmology}, 
(Harwood , Chur, Switzerland, 1990). 

\bibitem{Ve1} G. Veneziano, 
{\sl Strings, Cosmology, ... And A Particle }, 
CERN-TH.7502/94

\bibitem{Ve2} G. Veneziano, 
{\sl Scale Factor Duality For Classical And Quantum Strings}, 
Phys. Lett. {\bf B265}, (1991), 287. 

\bibitem{Ve3} G. Veneziano, 
{\sl The Graceful Exit Problem in String Cosmology}, 
Phys. Lett. {\bf B329}, (1994), 429. 

\bibitem{TV}A. A. Tseytlin and C. Vafa, 
{\sl Element of String Cosmology}, 
Nucl. Phys. {\bf B372}, (1992), 443. 

\bibitem{Ts2}A. A. Tseytlin, 
{\sl String Cosmology and Dilaton}, 
Eric. Theo. Phys. (1992), 202.  

\bibitem{Ts3}A. A. Tseytlin, 
{\sl Cosmological solutions with dilaton and maximally symmetric space in
 string theory}, 
Int. J. Mod. Phys. {\bf D1}, (1992), 223. 

\bibitem{Ts1}A. A. Tseytlin, 
{\sl Duality and Dilaton}, 
Mod. Phys. Lett. {\bf A6}, (1991), 1721.  

\bibitem{KY} K. Kikkawa and M. Yamasaki, 
{\sl Casimir Effect in Superstring Theories}, 
Phys. Lett. {\bf B149}, (1984), 367.

\bibitem{Sc} J. H. Schwarz, 
{\sl Superstring Compactification and Target Space Duality}, 
Strings: Stony Brook, (1991), 3, (hep-th/9108022). 

\bibitem{HV}G. t'Hooft and M. Veltman, 
{\sl One Loop Divergences in the Theory of Gravitation}, 
Ann. Inst. H. Poincare. {\bf A20}, 69, (1974). 

\bibitem{CD} S. Christensen and M. Duff, 
{\sl Quantum Gravity with A Cosmological Constant}, 
Nucl. Phys. {\bf 170B}, (1980), 480. 

\bibitem{BKK}  A. O. Barvinski, A. Kamenschik and B. Karmazin, 
{\sl The Renormalization Group for Non-Renormalizable Theories: Einstein
 gravity with A Scalar Field}, 
Phys. Rev. {\bf D48}, 3677, (1993). 

\bibitem{ST1} I. L. Shapiro and H. Takata, 
{\sl One Loop Renormalization of the Four-Dimensional Theory for Quantum
 Dilaton Gravity}, 
Phys. Rev.{\bf D52}, 2162, (1995).

\bibitem{ST2} I. L. Shapiro and H. Takata, 
{\sl Conformal Transformation in Gravity}, 
Phys. Lett. {\bf B361}, (1995), 31. 

\bibitem{IMM} T. Inagaki, S. Mukaigawa and T. Muta, 
{\sl A Soluble Model of Four-Fermion Interactions in de Sitter Space}, 
Phys. Rev. {\bf D52}, (1995), 4267. 

\bibitem{AIT} I. Antoniadis, J. Iliopoulos and T. N. Tomaras, 
{\sl One Loop Effective action around De Sitter Space }, 
(hep-th/9510112). 

\bibitem{ST3} I. L. Shapiro and H. Takata, 
in Preparation. 


\bibitem{Ab} L. F. Abbott, 
{\sl The Background Field Method Beyond One Loop}, 
Nucl. Phys. {\bf 185B}, (1981), 189.



\bibitem{De} B. S. DeWitt, 
{\sl Dynamical Theory of Groups and Fields}, 
(Gordon and Breach, NY, 1965). 

\end{thebibliography}
\end{document}